\documentclass[]{llncs}
\usepackage{llncsdoc,graphicx,algorithm}
\usepackage{float}
\usepackage[noend]{algorithmic}

\newcommand{\zzv}[1]{\vspace{#1}}
\newcommand{\zzi}{\itemsep-1.3mm\parsep-1.2mm}
\usepackage{pgfplots}
\usepackage{tikz}
\usepackage{subfigure}
\usepackage{color}
\usepackage{amsfonts}

\begin{document}
\newcounter{save}\setcounter{save}{\value{section}}
{\def\addtocontents#1#2{}%
\def\addcontentsline#1#2#3{}%
\def\markboth#1#2{}%
\title{Efficient Web-based Data Imputation with Graph Model}

\author{Yiwen Tang\inst{1} \and Hongzhi Wang\inst{1} \and Shiwei Zhang\inst{1} \and Huijun Zhang\inst{1} \and Ruoxi Shi\inst{1} 
\\ isabeltang147@gmail.com, wangzh@hit.edu.cn, ylxdzsw@gmail.com, zhjsss12@hotmail.com, shiruoxi@hit.edu.cn}
\institute{$^{1}$Harbin Institute Of Technology,China}

\maketitle

\begin{abstract}
A challenge for data imputation is the lack of knowledge. In this paper, we attempt to address this challenge by involving extra knowledge from web. To achieve high-performance web-based imputation, we use the dependency, i.e. FDs and CFDs, to impute as many as possible values automatically and fill in the other missing values with the minimal access of web, whose cost is relatively large. To make sufficient use of dependencies, We model the dependency set on the data as a graph and perform automatical imputation and keywords generation for web-based imputation based on such graph model. With the generated keywords, we design two algorithms to extract values for imputation from the search results. Extensive experimental results based on real-world data collections show that the proposed approach could impute missing values efficiently and effectively compared to existing approach.
\end{abstract}
\zzv{-3mm}
\section{Introduction}
\zzv{-3mm}
According to recent statistics, the quality of database degenerates over time and causes loss or even disasters\cite{ref1}. Data quality issues are to be solved to ensure the usability of data. For data quality issues, data incompleteness is one of the most pervasive data quality problems to handle\cite{ref2}. Due to its importance, data imputation has been widely studied \cite{ref3}\cite{ref4}. However, big data era brings new challenges for data imputation.

Firstly, for big data, knowledge is often insufficient for imputation, especially for the case with many missing values. Thus, extra knowledge is often required for big data imputation.

Second, even with sufficient knowledge, the accuracy of imputed value could not be ensured due to inconsistency and outdated data. For example, missing value could be filled according to existing knowledge base. With inconsistent or outdated data in corresponding item the knowledge base, the correct value could hardly be imputed.

Last but not least, efficiency is a non-negligible issue for big data imputation. Big data may contain many missing values to imputation. Timely imputation requires efficient imputation algorithms.

Facing to these challenges, some approaches have been proposed.
Web is often adopted as supplementary knowledge since web contains a large number of data sources. An example is WebPut~\cite{ref8}. However, WebPut requires many times of web search, whose cost is too large to meet the need of high efficiency.

Recently, crowdsourcing is very popular which organizations use contributions from Internet users to obtain needed services or ideas. However there has limitation on this method. Crowdsourcing allows anyone to participate, allowing for many unqualified participants and resulting in large quantities of unusable contributions which may reduce the accuracy.

Other approaches utilize initial data, in which approximate values are selected as the imputing values, according to the distribution characteristics or the constrains between attributes~\cite{ref10}\cite{ref3}\cite{ref11}. However, these methods are only suitable for numerical attributes but fail to impute category attributes.

In order to solve such problems, we introduce an optimized web-based data cleaning method. For both efficiency and effectiveness issues, we attempt to select proper missing values to be imputed according to the web based on the dependency between values, i.e functional dependency (FD) or conditional dependency (CFD), such that other missing values could be imputed accurately based on the imputed values. With the consideration of complex dependency relationships among missing values, we model such relationships as a directed graph, which is called statistical dependency graph (SDG). In a SDG, we introduce three kinds of nodes, attribute nodes, condition nodes and logic relation nodes to represent attributes, condition from CFD and the relationship among attributes, respectively. Since the value of an attribute may be implied from other values according to the relationship, it is necessary for us to add a logic node between attribute nodes.

Furthermore, to achieve high filling ratio according to web, we leverage the capabilities of web search engines towards the goal of completing missing attribute values based on the keyword group obtained from SDG. We input the keyword group in the search engine to get the text dependency from Internet. Text dependency is the relation between attribute and text or attribute and attribute, which is the pattern for data cleaning. For some data set, it is difficult to find pattern. In this case, we just use keyword group values for searching.

The contributions of this paper are summarized as follows.
\zzi
\begin{itemize}
\item We propose an optimized web-based big data imputation approach based on the dependency among missing values to increase the efficiency without the loss of effectiveness. As we know, this is the first work to combine the dependency with web-based imputation.

\item To increase the filling ratio, we use multiple search engines to cover as many data sources as possible. With the consideration of the variety in representation, we develop pattern discovery and keyword-group-based search algorithms to extract proper information for imputation from the search results.

\item We conduct extensive experiments to test the performance of proposed approaches. Experimental results show that our approach could impute large data sets efficiently and effectively for various data types.
\end{itemize}
\zzi
The remaining parts of this paper are organized as follows. Section 2 introduce background and overview the approach. We define graph model in Section 3. The imputation approach based on web search is proposed in Section 4. Experimental results and analysis are presented in Section 5. Section 7 draws the conclusions.

\zzv{-3mm}
\section{Overview}
\zzv{-3mm}
In this section, we introduce background and overview our approach.

\zzv{-3mm}
\subsection{Introduce To Functional Dependency}
\zzv{-2mm}
A functional dependency is a constraint that describes the relationship between attributes in a relation. A functional dependency $FD: X\to Y$ means that the values of $Y$ are determined by the values of $X$. Conditional functional dependency(CFD) \cite{ref9} is proposed as a novel extension of FDs. FD holds on all the tuples in the relation, while CFD is an FD which holds on the subset of tuples satisfying a certain condition. Compared with FDs, CFDs incorporate the bindings of semantically related values which can effectively capture the consistency of data. Thus, CFD represents the dependency relationship among attributes more subtly. As a result, We can use FDs or CFDs to find the missing data based on known values according to their dependencies.
\begin{table}[H]
\label{Fig.1}
\zzv{-2mm}
\caption{NBA Team Example}
\scriptsize
\begin{center}
\begin{tabular}{lllllll}
\hline\noalign{\smallskip}
ID &Team & Start-End & Arena & Location
  & Capacity & Coach\\
\noalign{\smallskip}
\hline
\noalign{\smallskip}
$t_{1}$& Golden State Warriors &1964-1966 &CivicAuditorium& SanFrancsicoCA & 7500 & \\
$t_{2}$& Golden State Warriors &1964-1966 &USFMemorialGym & & 6000 &A.Hannum \\
$t_{3}$&Oklahoma City Thunder& 2007-2014 &  & OklahomaCityOK &18203 & \\
$t_{4}$&               &1966-1967 & CivicAuditorium &  & & \\
$t_{5}$&Atlanta Hawks& 1949-1951 & WheatonFieldHouse & &  &Arnold Jacob\\
\hline
\end{tabular}
\zzv{-6mm}
\end{center}

\end{table}

We use an example to illustrate FD and CFD.
\zzi
\begin{example}
Consider Table 1, which specifies the NBA team VS arena in terms of the Team, Start-End, Arena, Location and Capacity.
A set of CFDs and FDs on this data set is as follows.

$f_{1}:[Arena]\to [Location, Capacity]$

$f_{2}:[Start-End, Arena]\to[Team, Location, Capacity] $

$f_{3}:[Start-End, Team]\to[Arena] $

$f_{4}:[Arena]\to[Team], 80\% $

$f_{5}:[Capacity]\to[Location], 70\% $

$f_{6}: [Coach= A.Hannum , Start-End] \to [Team]$

$f_{1}$ is an FD and shows that Arena determines the value of $Location$ and $Capacity$. In contrast, $f_{6}$ holds on the subset of tuples that satisfies the constraint ``$Coach = A.Hannum$'', rather than on the entire table. Compared with FDs, $f_{6}$ cannot be considered an FD since $f_{6}$ includes a constraint with data values in its specification. In a word, we can utilize the values of some attribute to determine missing values for imputation according to FDs, and CFDs can compensate FDs for achieving potential relations conditionally to find missing value in a domain.
\end{example}
\zzi
\zzv{-3mm}
\subsection{Overview Of Our Approach}
\zzv{-2mm}
For efficient and effective imputation, we develop the web-based imputation framework in Fig~\ref{Fig.3}. The basic flow of the framework is to construct some keyword groups according to the attribute names, values and dependencies, retrieve results from search engines with keyword groups and extract values for imputation from search results.

To reduce the number of values to be imputed according to web, we adopt dependencies. With the consideration of multiple dependencies on the data set, we design a graph model for dependencies, \emph{Statistical Dependency Graph}(SDG), which will be defined in Section~\ref{sec:model}.

As shown in Fig~\ref{Fig.3}, the first step constructs the SDG according to the dependencies among data. Through the statistics information, confidence, the statistics of the data set representing the possibility that a parent node determines a child node, is introduced as the weight of the SDG. Such confidences are used to generate the optimal keyword group for the further step. The details of SDG construction is described in Section 3.1. According to dependencies in SDG, we use naive Bayes~\cite{ref12} to impute missing values based on existing values directly without accessing web to accelerate imputation (Section 3.2). After such internal imputation step, we generate keyword groups by the largest confident single sink graph discovery algorithm to the target vertex on SDG (Section 3.3), where a single sink graph is a special subgraph of SDG with the sink as the attribute node corresponding to a missing value and the sources as the attribute with existing values in the original data.

The selected optimal keyword group is submitted to web search engines and the results are used for imputation. To obtain proper keywords for the search engine, we obtain the pattern though keywords group value from clean tuples. Otherwise, it means that the missing value could hardly be extracted with some fixed pattern. Thus, we submit the keywords group with the attribute name corresponding to the missing value to the engine and extract the the value for imputation with dictionary.
\zzi
\begin{figure}[t]
\centering
\includegraphics[width=7.5cm,height=7.5cm]{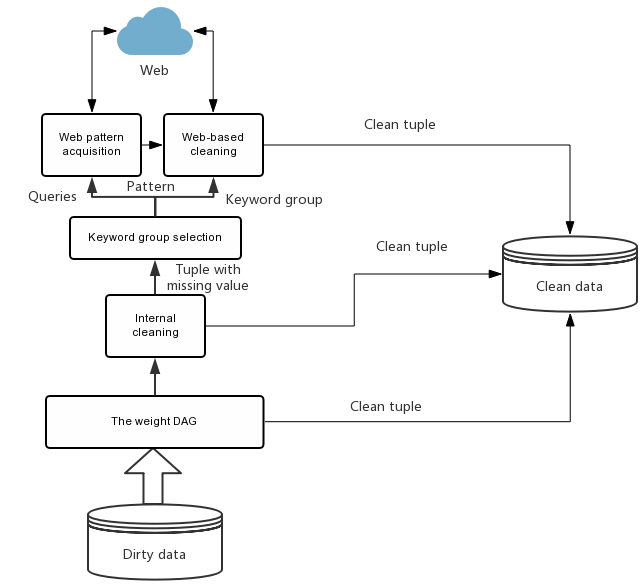}
\zzv{-2mm}
\caption{The Imputation Framework}
\zzv{-4mm}
\label{Fig.3}
\end{figure}

\zzv{-1mm}
We use an example to illustrate the flow.
\zzv{-3mm}

\begin{example}
We attempt to imputation missing values in the NBA data set in Table 1. According to our model, we firstly construct the SDG based on $FDs(f_{1},f_{2}, f_{3}, f_{4},
f_{5})$ and CFD($f_{6}$) shown in Figure 2. In this example, we assume that all of the weights of $f_{1}, f_{2}, f_{3}, f_{6}$ are 100\%, and the weights of $f_{4},f_{5}$ are 80\% and 70\%, respectively.

\begin{figure}[H]
\centering
\includegraphics[width=0.6\linewidth]{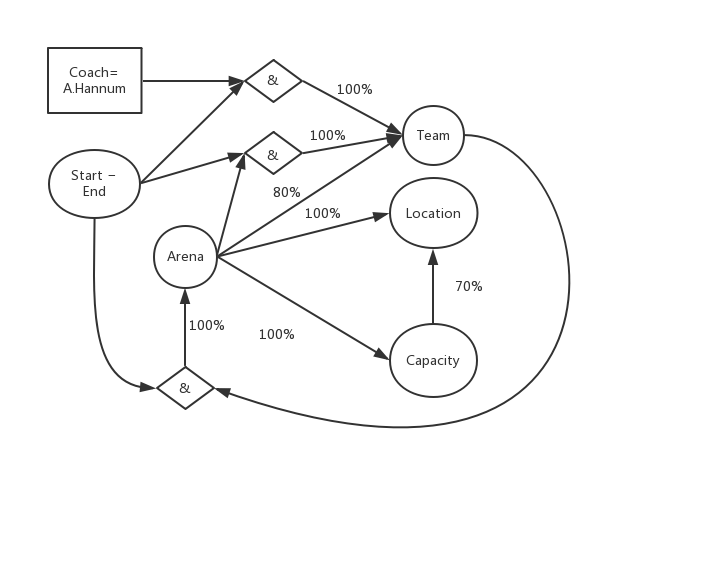}
\zzv{-15mm}
\caption{SDG based on Table 1}
\zzv{-6mm}
\label{Fig.15}
\end{figure}
Then, we proceed to internal imputation according to the SDG based on naive Bayes. In this example, we can impute the missing value of attribute $Capacity$ based on $f_{1}$. Firstly, we obtaion all possible values of attribute $Capacity$ such as 7500, 18203. According to $f_{1}$, we find that attribute $Arena$ determines attribute $Capacity$. Since the $Capacity$ of $CivicAuditorium$ is 7500 from $t_{1}$ in data set, we impute 7500 in $t_{4}$.

After the internal imputation, we obtain the keyword group from the single sink graph according to SDG. In the SDG, the single sink graph with maximum confidence for attribute $Arena$, $Location$, $Capacity$, $Team$ is shown in Fig.3(a), Fig.3(b), Fig.3(c), Fig.3(d), respectively. The confidence of the four graphs are all 100\% based on Fig.2. Since the vertex with short path to the sink have high confidence to imply the value corresponding to the sink, we obtain the single sink graph through breadth first search (BFS). The weight of a graph is the product of confidences of all edges in the graph. From the single sink graph, we generate four keyword groups $\{Start-End, Team, Arena\} \{Arena, Location\} \{Arena, Capacity\} \{A.Hannum, Start-End, Team\}$, respectively.

\zzi
\begin{figure}[H]
\centering
\subfigure[]{
\includegraphics[width=0.4\textwidth]{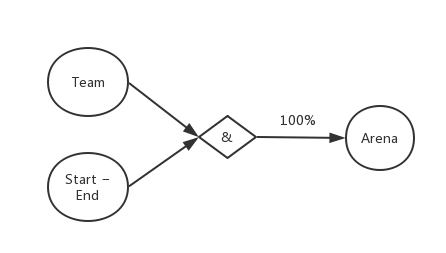}
}
\subfigure[]{
\includegraphics[width=0.4\textwidth]{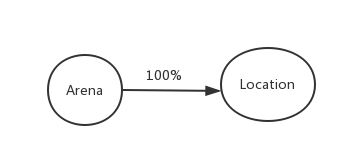}
}
\subfigure[]{
\includegraphics[width=0.4\textwidth]{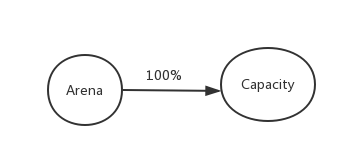}
}
\subfigure[]{
\includegraphics[width=0.4\textwidth]{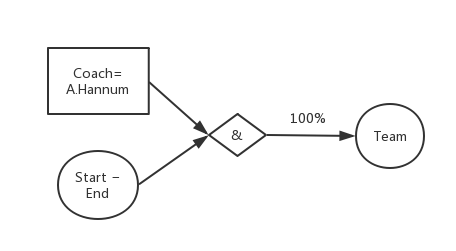}
}
\zzv{-2mm}
\caption{Single Sink Graph}
\zzv{-6mm}
\label{Fig.15}
\end{figure}

For example, if we want to find the missing value of $Location$ in $t_{5}$, We firstly submit keyword group $\{CivicAuditorium,SanFrancsicoCA\}$ to the search engine and discover the pattern from search result. The pattern is ``$A_{1} in A_{2}$'', where $A_{1}$ and $A_{2}$ are attributes and ``$in$'' is context.

Next, we search ``$WheatonFieldHouse$ in'' to obtain the missing value of $Location$ on web. We use a dictionary to extract the value which is the closest with keyword group. On the other hand, if we cannot achieve the pattern, then we search $\{WheatonFieldHouse,Location\}$ to find the missing value according to the optimal keyword group $\{Arena, Location\}$. The way of value extraction is based on dictionary as well. In the example, we obtain result \emph{Get information about WheatonFieldHouse in Wheaton, IL, including location, directions, reviews and photos \dots} from web. We match ``WheatonIL" from the result based on dictionary. Then we choose ``WheatonIL" as the value for imputation.
\end{example}

\zzv{-3mm}
\section{Graph-based Model}
\zzv{-3mm}

\label{sec:model}
In this section, we introduce graph-based model as well its usage in imputation. We first define the structure of SDG in Section 3.1 and introduce internal imputation based on naive Bayes in Section 3.2 according to SDG. As internal imputation cannot fill all missing values, we then generate keyword groups to be submitted to search engine for web-based imputation. Keyword group selection approach is proposed in Section 3.3.

\zzv{-3mm}
\subsection{SDG Definition}
\zzv{-2mm}
The goal of SDG is to capture the complex (conditional) dependency relationships among attributes for further internal imputation and keyword group generation. Since a dependency relationship between two attributes could be naturally modelled as an directed edge, we model the set of dependencies as a weighted directed graph. The SDG based for $f_1$ to $f_6$ is shown in Fig.2.

With such considerations, we define SDG as follows.
\zzi
\begin{definition}
A \emph{Statistics Dependency Graph}(SDG) is a weighted directed graph $G$=($V_{a}$, $V_{l}$, $V_{c}$, $E$, $W$) corresponding to a CFD set $S$, where $V_{a}$ is the set of attribute nodes, $V_{l}$ is the set of logic nodes, $V_{c}$ is the set of condition nodes, $E$ is the set of directed edges, and $W$ is the weight function of edges. $W$ is the confidence which is the ratio of number of tuples which satisfied the DFs and the total number of tuples. For each CFD $f\in S$, $f: \{C, a_{1}, a_{2}, \dots a_{n}\} \to \{d_{1}, d_{2}, \dots d_{m}, confidence\}$. $a_{1}, a_{2}, \dots a_{n}, d_{1}, d_{2}, \dots d_{m}$ are attributes which are represented as attribute nodes ($v_{a} \in V_{a}$). $C$ represents the condition, denoted as the condition node $v_{c} \in V_{c}$. When multiple attributes determine other attributes, we introduce a logic node $v_{l} \in V_{l}$ with the parents as  $a_{1}, a_{2}, \dots a_{n}$ and the children as $d_{1}, d_{2}, \dots d_{m}$. Confidence is represented as the weight on the graph of the graph.
\end{definition}
\zzi
The structure of SDG based on Example 1 is shown in Fig.2. According to $f_{1}-f_{6}$, there are five attributes in FDs and CFD. Additionally, in $f_{2}$, $f_{3}$ and $f_{6}$, multiple attributes determine other attributes. For example, when both attribute $Start-End$ and $Arena$ are known, we can determine attribute $Team$, $Location$ and $Capacity$ according to $f_{2}$. We can see that attribute $Start-End$ and $Arena$ has logical relationship. Also $f_{6}$ has a condition to restrain the dependency. $[Coach= A.Hannum , Start-End] \to [Team]$, which consists of a pattern tuple $( A.Hannum , \_ , \_  )$. The condition is that the value of an attribute $Coach$ is $A.Hannum$. As a result, we construct a SDG with five attribute nodes($V_{a}$), three logic nodes($V_{l}$) and a condition node($V_{c}$). We can obtain the weight of FDs and CFD according to the statistics of the data sets. In Example 1, the confidence of $f_{1}-f_{6}$ are 100\%, 100\%, 100\%, 80\%, 70\% and 100\%, respectively according to the weights of SDG.

From this example, we could discover that SDG could effectively represent the depending relationship among attributes according to FDs and CFDs with their confidences.

\zzv{-3mm}
\subsection{Internal Imputation}
\zzv{-2mm}
Considering the large cost of web accessing, we fill missing values according to the internal information of the data before web search. We choose naive Bayes approach in this step since it has several advantages. First of all, Bayes is both time and space efficient. We can look up all the probabilities with a single scan of the database and store them in a table. Besides, Bayes can handle both numerical values and categories.  Therefore, we use Naive Bayes for imputation according to SDG. That is to infer some missing values with existing values according to the FDs or CFDs.

We use Naive Bayes to impute data from the original data set. The approach has three stages, the preparation stage, classifier training stage, and application stage. In the first stage, we obtain the attribute set $A$ which can determine attribute of missing value. Then we obtain all possible values from other tuples as the candidate $d_{j}$. In the second stage, we calculate the probability $P$ for each candidate and the conditional probabilities for each attribute from $A$. In the final stage, $P(A|y_{i})P(y_{i})$ are calculated for each candidate and we select the candidate from the maximum term as the data for imputation. The definition is as follows.

Attribute $A_{1}, A_{2}, \dots, A_{n}$ determine attribute $D$ according to the dependency. There is a missing value in attribute $D$ in one tuple $x=\{a_{1j}, a_{2j}, \dots, a_{kj},\_, a_{k+1j}, \dots, a_{nj}\}$ of the data set($1\le k \le n-1$). We first scan the table to obtain all possible values of $D$ denoted by $d_{1}, d_{2}, \dots, d_{m}$.

Thus, for each $d_j$, we need to calculate the probability $P(d_j,  a_{1j}, a_{2j} \dots a_{nj})$. With the assumption that attributes are independent of each other, according to the Bias theorem\cite{ref12}, at first we compute $P(a_i|d_j)$ for each $a_i$ and $d_j$. Then for each  $d_j$, we calculate as follows.
\zzi
\begin{eqnarray}
\label{Bayes}
\zzv{-3mm}
P(d_{i}, a_{1j}, a_{2j}, \dots, a_{nj})= P(d_{i}) \times P(a_{1j} | d_{i})\times P(a_{2j} | d_{i}), \dots, P(a_{nj} | d_{i}) (1\le i\le m).
\zzv{-3mm}
\end{eqnarray}

Based on the deduction, we obtain the probability of all possible values of attribute $D$ in turn. The threshold $k$ is introduced to compare with the highest probability. If the highest probability reaches $k$, then the corresponding value will be filled in tuple $x$. Otherwise, we search on web. For a CFD like $f_{6}$, we obtaion the possible values from the tuple which satisfies the condition.

\zzv{-3mm}
\begin{algorithm}[H]
\caption{Internal Imputation based on Bayes}
\label{Fig.18}
\begin{algorithmic}[1]
\REQUIRE ~~\\
    data set, threshold $K$
\ENSURE ~~\\
    missing value
\FOR{$tuple(attr_{0}, \dots, attr_{n})$ in data set}
    \IF{$tuple (a_{0}, \dots., a_{n})$ has missing value of $attr_{i}$}
    	\STATE get $x=\{d_{1}\dots d_{m}\}$
   	\STATE /* where $d_{j}$ $(1\le j\le m)$ is each possible value of $attri_{i}$ */
    \FOR{each $d_{j}$$\in$$x$}
        \STATE $P(d_{j}, a_{0}, \dots,a_{i-1}, a_{i+1}\dots, a_{n}) = P( d_{j} ) \sqcap i P ( a_{i} | d_{j} )$
    \ENDFOR
    \ENDIF
\ENDFOR
\STATE select $(max P(d_{j}, a_{0}, \dots,a_{i-1}, a_{i+1}\dots, a_{n}), d_{j} )$
\IF{$d_{j} < k$}
	\STATE missing $value=NA$
\ELSE
	\STATE missing $value = d_{j}$
\ENDIF
\end{algorithmic}
\end{algorithm}	
\zzv{-3mm}

The pseudo code of internal imputation based on Bayes is shown in Algorithm 1. We obtain the possible values of missing attribute $d_{j}$ from other tuples (Line 3). According to \ref{Bayes}, we calculate the probability for each $d_{j}$ (Line 6). Then we select the value $d_{j}$ of maximum $P(d_{j}, a_{0}, \dots,a_{i-1}, a_{i+1}\dots, a_{n})$ (Line 7). If $d_{j}$ is less than the threshold $K$, we cannot impute with $d_{j}$ and give up the imputation (Line 8,9). Otherwise, we impute missing value of $d_{j}$(Line 11).

We use an example to illustrate this approach.
According to the NBA Team Example in Table 1, $t_{4}$ has missing values in attribute $Team$, $Location$, $Capacity$ and $Coach$. We can impute value of attribute $Location$ and $Capacity$ based on $f_{1}$. Firstly, the possible values of $Location$ are $\{SanFrancsicoCA, OklahomaCityOK\}$ and those of $Capacity$ are \{7500, 18203\}.
According to \ref{Bayes}, $P(SanFrancsicoCA, CivicAuditorium) = P(SanFrancsicoCA) \times P(CivicAuditorium|SanFrancsicoCA)$=1 which is larger than probability of $OklahomaCityOK$ (0), $P(7500, CivicAuditorium) = P(7500) \times P(CivicAuditorium|7500)$=1, which is larger than probability of $18203$(0). As a result, the value of Attribute $Location$ and $Capacity$ are $SanFrancsicoCA$ and 7500, respectively.

Note that all missing values cannot be imputed with internal information. For example, in $t_{5}$, the attribute $Location$ and $Capacity$ cannot be imputed according to other tuples. The probability of the candidate \{$SanFrancsicoCA, OklahomaCityOK$\} for $Location$ and \{$7500, 6000, 18203$\} for $Capacity$ are 0, respectively. For such cases, we have to obtain such information from external knowledge sources, i.e. web. Then, we discuss how to generate keyword groups for web-based imputation.

\zzv{-3mm}
\subsection{Keyword group selection}
\zzv{-3mm}
Since we attempt to use search engine for web-based imputation. To make sufficient and efficient use of search engine, proper keywords generation is crucial. In this section, we define single sink graph and discuss the generation approaches for proper keyword groups. The definition is as follows.

\begin{definition}
A \emph{Single sink graph} is a subgraph of SDG, $G* \subset G$. $G*$ has a sink and a set of source.
We define the attribute whose value $v_t$ needs to be imputed as the sink node $t$. The attributes corresponding to the existing values are the sources while each other node from a source $s$ to $t$ corresponds to an attributes with missing values.
\end{definition}

We find that one node may have many previous nodes in SDG. It means that many dependencies correspond to one attribute. If we randomly choose a single sink graph for data imputation, the accuracy is affected because the source node may not have the closest relation with the sink node. As a result, using such keyword to search may get the wrong value in higher probability. Therefore, it requires to choose a proper dependency for imputation.
As discussed in Section 3.1, a SDG denotes the relation among attributes and also reflects the dependency of the attributes.
We can find the most confident dependency  through SDG. In the optimal keyword group, the confidence as the production of all edges should be maximized. To this end, we develop an efficient algorithm for finding the optimal keyword group by breadth first search to select the single sink graph with maximum production of confidences in SDG.

In this algorithm, the confidence means how much the probability for a node to determine its child. Therefore, a higher confidence means higher dependency. For efficiency and accuracy, we should choose the single sink graph with highest confidence from SDG. Considering that the value of parent node may miss as well, we should continue to find the previous node of the parent node until we find the node with existing data as the source node. During BFS, we find that the dependency will be weak with continuous tracking back to the previous node. As a result, we define the production of confidences of all edges as the weight of the single sink graph. We start at an attribute with missing values. Then we find all its parent nodes and record the corresponding confidences. If the values of some parent nodes are missing, we continue to find the previous node and record the production of the confidence. We introduce a threshold $k$ to denote the credibility of the keyword group. If the cost is less than $k$, the keyword group should not be submitted to the search engine. Finally, we select the single sink graph with maximum weight and obtain the optimal keyword group which contains all nodes in the graph.

\zzv{-5mm}
\begin{algorithm}[H]
\caption{Optimal Keyword Group Selection}
\label{Fig.10}
\begin{algorithmic}[1]
\REQUIRE ~~\\
    missing value, SDG, threshold $K$
\ENSURE ~~\\
    $G*$
\STATE Initialize $CandidateRule = \varphi$
	\FOR{each $rule$ $\in$ $RuleList$}
	\FOR{each $condition$ $\in$ $rule.ConditionList$}
		\STATE search for the attribute in $condition	$
	\IF{meet $condition$}
		\FOR{each $Dependentproperties$ $\in$ $rule.DependentPropertiesList$}
			\STATE $c = rule_{c} \times \sqcap c(c \in \textit{condition of Dependent properties})$
			\STATE add $rule$ to $CandidateRule$
		\ENDFOR
	\ELSE
		\STATE add $rule$ to $CandidateRule$, $c = 0$
	\ENDIF
	\ENDFOR
\ENDFOR
\STATE $G* = maxrule _{c}$
\IF{$G*_{c} < K $then}
	\STATE result NA
\ENDIF
\end{algorithmic}
\end{algorithm}
\zzv{-5mm}

Algorithm 2 shows the optimal keyword group selection algorithm. For each $rule$ in $RuleList$, If meet $condition$ in correspondence $ConditionList$, we calculate the production of $confidence$ and add $rule$ to the $CandidateRule$(Line 2-8). Otherwise, we set confidence as 0(Line 9, 10). After that, we obtain the optimal $G*$ with the maximum confidence(Line 11). If the confidence is less than threshold $K$, the optimal $G*$ is abandoned(Line 12, 13).

We use an example to illustrate this algorithm.

As is shown in Fig.2, the SDG is based on Table 1. In $t_{5}$, the value of attribute \emph{Location} is missing. According to Fig.2, two attributes $Capacity$ and $Arena$ determine the attribute $Location$. Since the value of $Capacity$ is missing as well, we need to find its parent which is attribute $Arena$. The first single sink graph is $\{Arena\to Capacity\to Location\}$, with weight $W = 100\% \times 70\%$. It is obvious that we can determine the value of $Capacity$ according to attribute $Arena$, but the probability is only 70\% to determine value of $Location$ according to attribute $Capacity$. In a word , we may have probability 30\% to impute the wrong value. The second single sink graph is $\{Arena\to Location\}$. Compared with the first graph, the confidence is 100\% which outnumbers the first graph confidence. For accuracy, we define threshold $K=80\%$ to constrain the graph. In this situation, the first graph will be abandoned.

Keyword group selection is important for pattern mining and web searching. We can effectively find the missing value according to the keyword group based on the single sink graph with maximum confidence.

\noindent{\bfseries Time and Space Complexity.} In this algorithm, we scan all missing values. In this part, the time and space complexity is $O(n)$. For missing value, we obtain a single sink graph from SDG by BFD. As a result, the time complexity is $O(|V|+|E|)$ and the space complexity is $O(B)$, where $B$ is the maximum branch coefficient.

\zzv{-3mm}
\section{Data Imputation Based On Web}
\zzv{-3mm}

To find proper information, we submit generated keyword groups to search engine and extract values for imputation from returned results.
Most information on the web is text, and we require to extract proper value from text for imputation. For such extraction, we define the pattern in Section~\ref{sec:text}. We also propose the pattern mining approach and pattern-based imputation value extraction in Section~\ref{sec:mining}. For the special case without significant pattern, we also give the solution to achieve high filling ratio.
\zzv{-4mm}
\subsection{Text Dependency}
\zzv{-2mm}
\label{sec:text}
Intuitively, the values in the text with special context corresponding to some entities in a tuple. For instance, in the Table 1, when we search keywords $CivicAuditorium$ and $SanFrancsicoCA$, we will achieve the results that Civic Auditorium is a multi-purpose venue in San Francisco which reflect the relation between attribute $Arena$ and $Location$.
Thus, we attempt to use such context with variables representing values in the tuple as the pattern. With such pattern, we could extract the value for imputation according to existing values in the tuple. For example, we want to find the missing value in $t_{5}[Location]$ based on $f_{1}$. We firstly search $CivicAuditorium$ $SanFrancsicoCA$ and mine pattern from the results. The obtained pattern is ``$[Arena]$ in $[Loacation]$''. After that, we search ``$WheatonFieldHouse$ $in$'' and obtain result ``\emph{Get information about WheatonFieldHouse in Wheaton, IL, including location, directions, reviews and photos \dots} ''. From this phrase, we know that \emph{Wheaton} is in the position of $[Loacation]$. We match it with dictionary and find the imputed value``$WheatonIL$''.

Motivated by this, we define text dependency as the pattern as follows.

\begin{definition}
A pattern is in form of $P = X_{1}A_{1}X_{2}A_{2}\dots X_{n}A_{k}X_{k+1}$, where $X_i$ represents a phrase and $A_i$ is an attribute in the table $T$ to impute. For a tuple $t\in T$, we denote ``$X_{1}t[A_{1}]X_{2}t[A_{2}]\dots X_{n}t[A_{k}]X_{k+1}$'' as $P(t)$.
For a text set $\mathbb{S}$, a text dependency $P$ and a table $T$, $S_{P, T}$=\{$S|S\in \mathbb{S}\wedge\exists t\in T$, $P(t)\in S$\}.
For a pattern $P$, if $|S(P, T)|/|\mathbb{S}|\geq Q$, where $Q$ is a threshold, $P$ is called a text dependency on $\mathbb{S}$.
\end{definition}

For example, a text dependency could be ``$WheatonFieldHouse in$'' as is mentioned above. With such patter, $WheatonIL$ could be extracted from the text in the return result according to the value $CivicAuditorium$ and $SanFrancsicoCA$. Such value could be imputed to the value of attribute $Location$ in $t_{5}$ in Table 1.
\zzv{-4mm}
\subsection{Pattern mining}
\zzv{-2mm}
\label{sec:mining}

We observe that only the pattern which reflects the relation between attributes are useful for us to obtain result such as verbs and prepositions.
However, sometimes we may extract some meaningless word which cannot help us to achieve correspondence text.
Thus, according to the keyword group, we need to mine the text which introduces the relation between attributes so that we can utilize the relation to depict the value we want. As a result, our algorithm search values of keyword group to retrieve result from web which contains such relation. If the number of phrase $S$ appearances exceeds a threshold $Q$, it is the pattern. We store all possible patterns.

During pattern mining, firstly, we use correct tuples to generate keywords. Then we submit the keyword to search engine and obtain the text. We mine the text dependency from results based on hash table storing the possible patterns. Finally we choose the phrase $S$ whose appearance number exceeds threshold $Q$ as the pattern.

\begin{algorithm}[t]
\caption{Pattern Mining Algorithm}
\label{Fig.9}
\begin{algorithmic}[1]
\REQUIRE ~~\\
    data set, $Tuples$, Threshold $Q$
\ENSURE ~~\\
    patterns
\STATE Initialible $PatternHash Table = \varphi$
\FOR{each $tuple$ in $tuples$}
    \STATE Initialible $TupleContentHash = \varphi$
    \STATE retrieve $Queries$ and get $Results$
    \FOR{each $string$ in $Results$}
        \IF{$string$ in $TupleContentHash$}
            \STATE get $start$ and $end$
            \FOR{each $substr$ includes $[start\dots end]$}
                \STATE replace$(substr value, flag)$
                \STATE add$(PatternHashTable, substr)$
            \ENDFOR
        \ENDIF
    \ENDFOR
\ENDFOR
\IF{ $|substr| < Q$}
\STATE abandon substr
\ENDIF
\RETURN $PatternHashTable.substr$
\end{algorithmic}

\end{algorithm}	

The pseudo code is shown in Algorithm 3. At first, we input correct tuples on web and obtain get the results(Line 2-4). Then we find the position of the attribute values in tuples (Line 5-7). In Line 8-10, we get the substring between attribute values and add it into the table. After that, We compare the appearance number of substrings in table with threshold $Q$ and obtaion the patterns which number exceeds the threshold $Q$ (Line 11-13).
We use an example to illustrate this algorithm.

\begin{example}
Consider a table which contains university names and principal names. To get the pattern, we select all the correct tuples on web i.e. ``\emph{Harbin Institute Of technology + YuZhou}'', ``\emph{Tsinghua  University + YongQiu}'',\emph{``Peking University + Enge Wang}''. We obtain the results from he search engine. The results are  \emph{ `` \dots YuZhou is the principal of Harbin Institute Of technology \dots}'', \emph{ ``\dots YongQiu is the present principal of Tsinghua University \dots}'', \emph{ ``Enge Wang served as the principal of Peking University }''. Then we extract the text which the first word and last word are from keyword and we insert every string in the text into the $PatternHashTable$. Suppose that the threshold $Q$ is 2. We compare all the values in the table with threshold $Q$ and we can get the patterns ``$[attribute_{1}]\textit{ the principal of }[attribute_{2}]$''.
\end{example}

We now analyze the complexity of Algorithm 2. We denote the number of the tuples after sample as $N$ , the number of attributes in each tuple as $M$ and the time spent in each web query as $T$, the time complexity of the algorithm is $N(M_{2}-M)T/2$, that is, $O(NM_{2}T)$.

\begin{example}
As is shown in Table \ref{Fig.6}, the process of mining text dependency in the form is as follows. First, all tuples in the form as follows are searched on web respectively and top 100 records are to be mined. In this example, we set threshold $Q=50$. Then we can obtain the occurrence number of phrase $S=[A_{1}]$ director $[A_{2}]$, $T>Q$ which result in that attribute $A_{1}$, and $A_{2}$ has text dependency.
\end{example}
\begin{table}[t]
\scriptsize
\caption{An Instance Of Mining Text Dependency}
\zzv{-6mm}
\label{Fig.6}
\begin{center}
\begin{tabular}{llllll}
\hline\noalign{\smallskip}
Num & Film & Director\\
\noalign{\smallskip}
\hline
\noalign{\smallskip}
1 & The Shawshank Redemption & Frank Darabont\\
2 & The Godfather & Francis Ford Coppola  \\
3 & Pulp Fiction & Quentin Tarantino  \\
4 & Schindler's List & Steven Allan Spielberg  \\
5 & Fight Club & David Fincher   \\
6 & One Flew Over the Cuckoo's Nest & Milos Forman  \\
7 & Inception & Christopher Nolan  \\
\hline
\end{tabular}
\end{center}
\end{table}

\zzv{-4mm}
\begin{definition}
Data cleaning based on web. According to the information retrieved on web, we mine text dependency of attribute. Then we utilize the text dependency to clean tuple.
\end{definition}
\zzv{-7mm}

\begin{table}[H]
\caption{an instance of data cleaning based on web}
\zzv{-6mm}
\scriptsize
\label{Fig.7}
\begin{center}
\begin{tabular}{llllll}
\hline\noalign{\smallskip}
Num & Film & Director\\
\noalign{\smallskip}
\hline
\noalign{\smallskip}
1 & The Shawshank Redemption & Frank Darabont\\
2 & The Godfather & Francis Ford Coppola  \\
3 & Pulp Fiction & Quentin Tarantino  \\
4 & Schindler's List & Steven Allan Spielberg  \\
5 & Fight Club & David Fincher   \\
6 & One Flew Over the Cuckoo's Nest & Milos Forman  \\
7 & Inception & Christopher Nolan  \\
8 & Se7en & \\
\hline
\end{tabular}
\end{center}
\zzv{-6mm}
\end{table}

\begin{example}
As is shown in Table \ref{Fig.7}, top seven tuples are correct and the value of the third attribute of the eighth tuple is missing. We can get the text dependency $S=[A_{1}]$ director $[A_{2}]$ by searching on the internet according to the top seven tuples. Then we retrieve $Query = Se7en$ director based on text dependency $S$ and the known value of the eighth tuple. Through analysis on the retrieved content, we can get $A_{2}= ``\textit{David Finch} "$.
\end{example}

\zzv{-4mm}
\subsection{Keyword-group-based Search}
\zzv{-2mm}

In practice, the pattern could hardly be obtained. To handle such case, we develop keyword-group-based search algorithm.

In practice, we may find that we cannot extract the pattern for some data set. In a word, data cleaning based on pattern may not suitable for all data sets. To this end, we develop an efficient method to solve this problem.

In this method, we firstly obtain the optimal keyword group and search it on web. Since the attributes from the single sink graph with the largest confidence has the closest relation with the attribute of missing value, we input source node values from the graph and the attribute of the missing value (sink node) to the search engine. We can get the value from the return result.

The pseudo code is shown in Algorithm 4. We first construct the keyword group which is ``$source node value + sink node from G*$'' (Line 1). Then we construct URL and request for website.(Line 2). After that, we extract words from the results with the dictionary(Line 3). Finally, we compare the average distance from keyword and select the minimum for data imputation(Line 4-6).

For example, according to the Table 1, we want to find the value of attribute $Team$ in $t_{4}$. We generate the optimal keyword group which is $\{Start-End,Arena,Team\}$. Therefore, we use $\{1966-1967,CivicAuditorium,Team\}$ as the keyword to search on web. After searching on web, we extract words in the return result by dictionary. We select the words with closest distance with the keyword in dictionary. Note that the dictionary is built by extracting all possible values in the column from web and existing values in the table.

\zzi
\zzv{-5mm}
\begin{algorithm}[H]
\caption{Keyword-group-based Search}
\label{Fig.18}
\begin{algorithmic}[1]
\REQUIRE ~~\\
    $G^{\star}$, $D$, source node value
\ENSURE ~~\\
    missing value $w$
\STATE Initialize $\textit{ k = source node value + sink node from }G^{\star}$
\STATE $T$ = getPage($K$)
\STATE extract words($T$,$D$)
\STATE avgDistance($w$, $K$, $T$)
\STATE select min(avgDistance)
\RETURN missing value $w$
\end{algorithmic}
\zzi
\end{algorithm}
\zzv{-5mm}

\section{Experimental Study}
\zzv{-2mm}

To test the performance of the proposed approach, we conduct extensive experiments in this section.

We implement all proposed approaches with Python and use baidu search engine. We run our experiments on a PC with an Intel Core i5-3470 CPU and 8GB memory, running 64bit Ubuntu 12.04. We compare state-of-art web-based data cleaning method \cite{ref8}.

To test the proposed methods comprehensively, we use following data sets.

The data sets above are complete relational tables. To test the performance of imputation, we omit values at random from the data sets and keep key attribute value in each tuple according to the FDs. We ensure that the previous node of missing value is imputed at least. Each proposed result is the average of 5 evaluations, that is, for each missing value percentage (1,5,20,30,40,50 and 60\%) refer to other experiments in web-based cleaning. In each evaluation, we remove data at random positions which means that we generate 5 incomplete tables. We then impute these incomplete tables using our model and evaluate the performance.
\zzv{-0mm}
\begin{table}[H]
\centering
\scriptsize
\zzv{-2mm}
\caption{Experimental Results On Accuracy}
\zzv{-1mm}
\subtable[The Accuracy Of Disney Dataset]{
      \begin{tabular}{llllllll}
\hline\noalign{\smallskip}
Missing ratio& 1& 2& 3& 4& 5& average \\
\noalign{\smallskip}
\hline
\noalign{\smallskip}
5	&1	&1   &0.75	&1	&1	&0.95\\
10	&1	&1   &1	&1	&0.83	&0.97\\
20	&0.91	&0.89&0.92	&0.89	&1   &0.92\\
30	&0.87	&0.93	&0.95	&1	&0.92	&0.93\\
40	&1	&0.88	&0.95	&1	&0.78	&0.92\\
50	&1	&0.82	&0.85&0.89	&0.92	&0.90\\
60	&0.93&0.91& 0.89&0.8	&0.96&0.90\\
\hline
\end{tabular}
       \label{Fig.13}
}
\qquad
\subtable[The Accuracy Of University Dataset]{
       \begin{tabular}{llllllll}
\hline\noalign{\smallskip}
Missing ratio& 1& 2& 3& 4& 5& average \\
\noalign{\smallskip}
\hline
\noalign{\smallskip}
5	&1	&1   &0.75	&1	&1	&0.95\\
10	&1	&1   &1	&1	&0.83	&0.97\\
20	&0.91	&0.89&0.92	&0.89	&1   &0.92\\
30	&0.87	&0.93	&0.95	&1	&0.92	&0.93\\
40	&1	&0.88	&0.95	&1	&0.78	&0.92\\
50	&1	&0.82	&0.85&0.89	&0.92	&0.90\\
60	&0.93&0.91& 0.89&0.8	&0.96&0.90\\
\hline
\end{tabular}
       \label{Fig.14}
}
\end{table}
\zzv{-1mm}
\begin{figure}[t]
\centering
\subfigure[Disney]{
\begin{tikzpicture}
 \begin{axis}[width=5cm, height=3.5cm, tick align=inside,xlabel = {$Missing Ratio[\%]$},
    ylabel = {$Accurarcy$},xtick={0,10,20,30,40,50,60},
    ytick={0.88,0.90,0.92,0.94,0.96,0.98,1},xmin=0, xmax=60,
    ymin=0.88, ymax=1,]
   \addplot[draw=blue,mark=square,] coordinates {(0,1)(5,0.95)(10,0.97)(20,0.92)(30,0.93)(40,0.92)(50,0.90)(60,0.90)};
 \end{axis}
 \end{tikzpicture}
}
\subfigure[University]{
\begin{tikzpicture}
 \begin{axis}[width=5cm, height=3.5cm, tick align=inside,xlabel = {$Missing Ratio[\%]$},
    ylabel = {$Accurarcy$},xtick={0,5,10,20,30,40,50,60},xmin=0, xmax=60,
    ymin=0.76, ymax=1,]
   \addplot[draw=blue,mark=square,] coordinates {(0,1)(5,0.96)(10,0.93)(20,0.87)(30,0.80)(40,0.80)(50,0.78)(60,0.81)};
 \end{axis}
 \end{tikzpicture}
}
\zzv{-3mm}
\caption{The Accuracy Of Web-based Data Imputation With Graph Model}
\zzv{-4mm}
\label{Fig.11}
\end{figure}
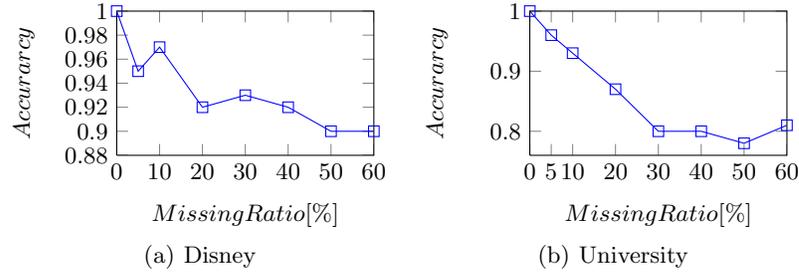

\zzv{-1mm}
\begin{enumerate}
\item Multilingual Disney Cartoon Table (Disney)\cite{ref8}: This table contains names of 51 classical Disney cartoons in 8 different languages collected from Wikipedia.

\item University Principle Information Table (Principle): This table contains 100 Chinese university and information of university includes address, city and principal which are collected from Wikipedia.
\end{enumerate}

\zzv{-5mm}
\begin{figure}[t]
\centering
\subfigure[Disney]{
\begin{tikzpicture}
 \begin{axis}[width=5cm, height=3.5cm, tick align=inside,xlabel = {$Missing Ratio[\%]$},
    ylabel = {$Time[s]$},xtick={0,5,10,20,30,40,50,60},
    ytick={0,5,10,15,20,25,30},xmin=0, xmax=60,
    ymin=0, ymax=30,]
\addplot[draw=blue,mark=square,] coordinates {(0,1)(5,7.1909838)(10,10.038072)(20,15.2028308)(30,15.6599458)(40,21.1689608)(50,24.3830968)(60,28.9417326)};
 \end{axis}
 \end{tikzpicture}
}
\subfigure[University]{
\begin{tikzpicture}
 \begin{axis}[width=5cm, height=3.5cm, tick align=inside,xlabel = {$Missing Ratio[\%]$},
    ylabel = {$Time[s]$},xtick={0,5,10,20,30,40,50,60},xmin=0, xmax=60,
    ymin=0, ymax=80,]
   \addplot[draw=blue,mark=square,] coordinates {(0,1)(5,10.080214)(10,20.768768)(20,27.723283)(30,39.085069)(40,45.246878)(50,61.359339)(60,75.516074)};
 \end{axis}
 \end{tikzpicture}
}
\zzv{-4mm}
\caption{The Time Cost Of Web-based Data Imputation With Graph Model}
\zzv{-4mm}
\label{Fig.20}
\end{figure}
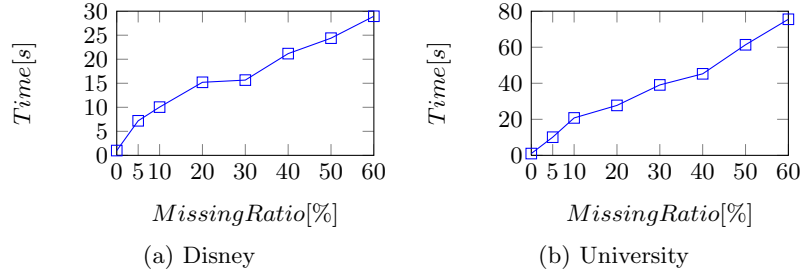

In this experiment, we first evaluate the accuracy of our model over two data sets. We propose tables of two datasets which contains specific results of 5 evaluations. As is shown in Fig.4(a) and Fig.4(b), the accuracy is pretty high and because of the random removal of value, there is no obvious relation between missing ratio and accuracy. That is good for cleaning big data in real world because our model ensure high accuracy of data sets regardless of the size of the data. The run time on these two data sets are shown in Fig.5(a) and Fig.(b). We observe that the time cost is nearly linear with the missing ratio. As mentioned above, we introduce threshold $k$ to restrain the data imputation for accuracy. Since there has 100\% dependency relation, the threshold $k$ does not affect the result in the two data sets according to our experiment. 

\zzv{-9mm}
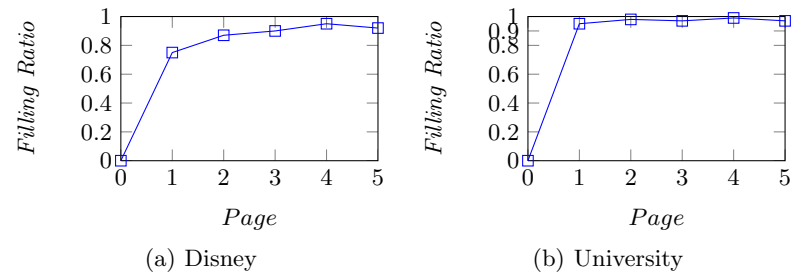
\begin{figure}[H]
\centering
\subfigure[Disney]{
\begin{tikzpicture}
 \begin{axis}[width=5cm, height=3.5cm, tick align=inside,xlabel = {$Page$},
    ylabel = {$\textit{Filling Ratio}$},xtick={0,1,2,3,4,5},
ytick={0,0.2,0.4,0.6,0.8,1},
    xmin=0, xmax=5,
ymin=0, ymax=1,
]
\addplot[draw=blue,mark=square,] coordinates {(0,0)(1,0.75)(2,0.87)(3,0.9)(4,0.95)(5,0.92)(6,0.93)};
 \end{axis}
 \end{tikzpicture}
}
\subfigure[University]{
\begin{tikzpicture}
 \begin{axis}[width=5cm, height=3.5cm, tick align=inside,xlabel = {$Page$},
    ylabel = {$\textit{Filling Ratio}$},xtick={0,1,2,3,4,5},
ytick={0,0.2,0.4,0.6,0.8,0.9,1},
    xmin=0, xmax=5,
ymin=0, ymax=1,
]
   \addplot[draw=blue,mark=square,] coordinates {(0,0)(1,0.95)(2,0.98)(3,0.97)(4,0.99)(5,0.97)(6,0.93)};
 \end{axis}
 \end{tikzpicture}
}
\zzv{-4mm}
\caption{The Filing Ratio Of Web-based Data Imputation With Graph Model}
\zzv{-4mm}
\label{Fig.22}
\end{figure}
\zzv{-9mm}
\begin{figure}[H]
\centering
\begin{tikzpicture}
 \begin{axis}[width=6.5cm, height=3.5cm, tick align=inside,xlabel = {$Missing Ratio[\%]$},
    ylabel = {$Accurarcy$},
    xmin=0, xmax=70,
    ymin=0.30, ymax=1,ybar interval=0.7]
   \addplot coordinates {(5,1)(10,0.97)(20,0.92)(30,0.93)(40,0.92)(50,0.90)(60,0.90)(70,0.90)};
\addplot coordinates {(5,0.8)(10,0.47)(20,0.40)(30,0.38)(40,0.35)(50,0.34)(60,0.33)(70,0.90)};
\legend{Our Model,Greedylter}
 \end{axis}
 \end{tikzpicture}
\zzv{-4mm}
\caption{Comparing The Accuracy Of Our Model And Greedy Iterative}
\zzv{-7mm}
\label{Fig.12}
\end{figure}
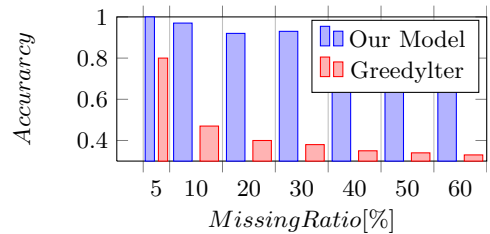

We also test the relation between $Filling Ratio$ and number of pages. Since search engines always the most important information on the first few pages, we observe that the $Filling Ratio$ is quite high when getting one page and increase gradually until the $Filling Ratio$ almost levels up. The experiments with $Disney$ and $University$ are shown in Fig.6(a) and Fig.6(b), respectively.

Finally, We compare the performance of our model and Greedy Iterative as is shown in Fig.7. We find that the accuracy of GreedyIter is better than One-Pass and Iterative, so we only propose the comparison with Greedy Iterative. It is clear that our approach reaches better accuracy at any missing ratio. The accuracy of Greedy Iterative drops with the increase of missing ratio. By comparison, our approach is not affected by the missing ratio.

\zzv{-3mm}
\section{Conclusion}
\zzv{-3mm}
We present the web-based model for processing dirty data effectively and efficiently. We have developed SDG to express FDs and CFDs. Our model has two parts for data cleaning. First we process dirty data by internal cleaning which can reduce waste time on web searching. We find that sometimes we can use known value to find missing value according to the sematic relations between the attributes. Then we can find the optimal keyword group from SDG which help us to find the missing value with high accuracy on web. The experiments show that our model can impute missing values more accurately for a relational datasets. In our future work, we will extend our model to deal with multiple kinds of datasets.

\zzv{-3mm}
\paragraph{Acknowledgement}
We thanks Dr. Zhixu Li for providing the data set of Disney.
\zzv{-3mm}
\bibliographystyle{abbrev}

\end{document}